\documentclass[preprint,aps]{revtex4}

\usepackage{graphicx}

\newcommand{\bq}{\begin{equation}}
\newcommand{\eq}{\end{equation}}
\newcommand{\bqn}{\begin{eqnarray}}
\newcommand{\eqn}{\end{eqnarray}}

\newcommand{\lb}{\label}
\newcommand{\noi}{\noindent}

\newcommand{\myfigure}[3]{
        \begin{figure}
        \centerline{
        \includegraphics{#1.ps}}
        \caption{#2}
        \label{#3}
        \end{figure}
}

\begin{document}

\title{On Anisotropic Dark Energy}

\author{R. Chan}
\email{chan@on.br}
\affiliation{Coordena\c c\~ao de Astronomia e Astrof\'{\i}sica,
Observat\'orio Nacional, Rua General Jos\'e Cristino 77, S\~ao
Crist\'ov\~ao, CEP 20921-400, Rio de Janeiro, RJ, Brazil}

\author{M. F. A. da Silva}
\email{mfasnic@gmail.com}
\affiliation{Departamento de F\' {\i}sica Te\' orica, Universidade do
Estado do Rio de Janeiro,
Rua S\~ ao Francisco Xavier $524$, Maracan\~ a, 
CEP 20550-013, Rio de Janeiro, RJ, Brazil}

\author{Jaime F. Villas da Rocha}
\email{jaime@mast.br}
\affiliation{Coordenadoria de Educa\c c\~ao em Ci\^encias, Museu de 
Astronomia e Ci\^encias Afins,
Rua General Bruce 586, S\~ao Crist\'ov\~ao, 
CEP 20921-030, Rio de Janeiro, RJ, Brazil}

\date{\today}

\begin{abstract}
Since the discovery of the accelerated expansion of the universe, it was
necessary to introduce a new component of matter distribution called dark
energy.  The standard cosmological model considers isotropy of the pressure
and assumes an equation of state $p=\omega \rho$, relating the pressure
$p$ and the energy density $\rho$.  The interval of the
parameter $\omega$ defines the kind of matter of the universe, related 
to the fulfillment, or not, of the energy conditions of the fluid.
The recent interest in this kind of fluid with anisotropic pressure,
in the scenario of the gravitational collapse and star formation, imposes
a carefull analysis of the energy conditions and the role of the components
of the pressure.  Here, in this work, we show an example where the
classification of dark energy for isotropic pressure fluids is used
incorrectly for anisotropic fluids. The correct classification and its
consequences are presented.
\end{abstract}

\maketitle

\section{Introduction}

Over the past decade, one of the most remarkable discoveries is
that our universe is currently accelerating. This was first
observed from high red shift supernova Ia \cite{agr98}-\cite{Obs5}, and
confirmed later by cross checks from the cosmic microwave
background radiation \cite{wmap}-\cite{wmap1} and large scale structure
\cite{sdss}-\cite{sdss5}.

In Einstein's general relativity, in order to have such an
acceleration, one needs to introduce a component to the matter
distribution of the universe with a large negative pressure. This
component is usually referred as  dark energy. Astronomical
observations indicate that our universe is flat and currently
consists of approximately $2/3$ dark energy and $1/3$ dark
matter. The nature of dark energy as well as dark matter is
unknown, and many radically different models have been proposed,
such as, a tiny positive cosmological constant, quintessence \cite{quint1}-\cite
{quint3},
DGP branes \cite{DGP1}-\cite{DGP2}, the non-linear F(R) models \cite{FR1}-\cite{
FR3},
and dark energy in brane worlds, among many others \cite{other1}-\cite{Brandt07}
; see also the review articles \cite{DE1}-\cite{DE2}, and references therein.

As mentioned before, the existence of dark energy fluids comes from the 
observations of the
accelerated expansion of the Universe and the isotropic pressure cosmological
models give the best fitting of the observations.  
Although some authors \cite{Koivisto} have suggested cosmological model with 
anisotropic and viscous dark energy in order to explain an anomalous 
cosmological observation in the cosmic microwave background (CMB) at the 
largest angles.
On the other hand, another very important issue in gravitational
physics is black holes and their formation in our universe. Although
it is generally believed that on scales much smaller than the horizon
size the fluctuations of dark energy itself are unimportant \cite{Ma99},
their effects on the evolution of matter overdensities may be
significant \cite{FJ97}-\cite{FJ97a}. Then, a natural question is how dark
energy affects the process of the gravitational collapse of a
star. It is known that dark energy exerts a repulsive force
on its surrounding, and this repulsive force may
prevent the star from collapse. Indeed, there are speculations that
a massive star doesn't simply collapse to form a black hole,
instead,  to the formation of stars that contain dark energy.
In a recent work, Mazur and Mottola \cite{Mazur02} have suggested a solution
with a final configuration without neither singularities nor horizons,
which they called "gravastar" (gravitational vacuum star).  In this case,
the gravastar is a system characterized by a thin (but not infinitesimal) shell 
made of stiff matter, which separates an inner region with
de Sitter spacetime from the Schwarzschild exterior spacetime.  
The elimination of the apparent horizon is done using suitable choice
of the inner and outer radius of the thin shell, in such way that the inner
radius be shorter than the horizon radius of de Sitter spacetime and the
outer radius longer than the Schwarzschild horizon radius.
In a later
work, Visser and Wiltshire \cite{Visser04} have shown that the gravastar is
dynamically stable.  
The possibility of the existence of objects like the
gravastar brings all the discussions about the fact that is unavoidable that 
gravitational collapse always forms a black hole.
As a result, black holes may not exist at all \cite{DEStar}-\cite{DEStar1}.
However, recently we have shown that although the possibility of the existence of gravastars cannot 
be excluded from such dynamical models, even if gravastars 
do indeed exist, this does not exclude the possibility of the existence of black holes
\cite{Chan08a}.

Another related issue is that how dark energy affects
already-formed black hole is related to the fact that 
 it was shown that the mass of a black hole decreases due
to phantom energy accretion and tends to zero when the Big Rip
approaches \cite{BDE04}-\cite{BDE04a}.
Gravitational collapse and formation of black holes in the presence of dark
energy were first considered by several works \cite{BHDE1}-\cite{BHDE4}.

Based on the discussions about the gravastar picture some authors have proposed
alternative models. Among them, we can find a Chaplygin dark star \cite{Paramos},
a gravastar supported by non-linear electrodynamics \cite{Lobo07},
a gravastar with continuous anisotropic pressure \cite{CattoenVisser05}. Beside these
ones,
Lobo \cite{Lobo} has studied two models for a dark energy fluid. One of them
describes a homogeneous energy density and another one which uses an
ad-hoc monotonic decreasing energy density, both of them with anisotropic
pressure.  In order to match an exterior Schwarzschild spacetime he has
introduced a thin shell between the interior and the exterior spacetimes.
In a recent work \cite{Chan08} we constructed another alternative model to 
black holes considering the possibility of gravitational trapping of dark
energy by standard energy, i.e., not dark energy.  
We have found that, in order to have static solutions, at least one of the
regions must be constituted by dark energy.
Then, in this context we could
have the dark energy sustaining the collapse of the standard energy, while the
standard matter would trap the dark energy.

Our interest in anisotropic dark energy comes from the fact that
some theoretical works on more
realistic stellar models suggest that superdense matter may be anisotropic at least in
certain ranges \cite{Ruderman}\cite{Canuto}. Different scenarios have been proposed which could give rise to
local anisotropy, e.g. pion condensation \cite{Hartle}, different kinds of phase transitions \cite{Itoh}\cite{Sokolov},
boson stars \cite{Ruffini}\cite{Gleiser}, type P superfluids, crystallization in white dwarfs \cite{Kippenhahn} among others
\cite{Letelier}\cite{Bayin}. Also in low density systems, local anisotropy may develop from the appearance
of an anisotropic velocity distribution, e.g. in the galactic halos of fermionic dark
matter \cite{Ralston}\cite{Madsen}. In the framework of general relativity, since the paper by Bowers and Liang \cite{Bowers},
the assumption of local anisotropy has been extensively studied (for further
references see \cite{Herrera}-\cite{Abreu}).
More recently, in the gravastar scenario, it was shown by Cattoen, 
Faber and Visser \cite{Cattoen} that these models must exhibit anisotropic 
pressures 
at least in their "crust" in order to be finite-sized objects. 
Besides, although the large scale distribution of dark energy is supposed to be
isotropic, as well as the standard  matter, we believe that it is reasonable to
admit that dark energy may be anisotropic at small scales, as well as the 
standard matter.

In this work we will demonstrate that the definition of anisotropic dark energy and 
phantom fluids can be more complex than the case of isotropic fluids, in terms of the energy conditions, 
since each pressure component can contribute in a different way to the convergence property of the gravity. 
Then, it is important to pay attention on this subject for systems
where anisotropy is relevant, as in the gravitational collapse or star formation \cite{Chan08}.

The paper is organized as follows. In Section
II we show the definitions of dark energy and phantom fluids. In Section III 
we present an example of misleading classification of dark energy and
phantom fluid and we also discuss some consequences.  Finally, in Section III we present our final considerations.

\section{Dark Energy and Phantom Fluid Definitions}

Dark energy and phantom fluids are characterized by the violation of certain energy conditions. 
These can be grouped in three kinds, that are: (i) weak energy conditions; (ii) dominant energy conditions; 
(iii) strong energy condition. In order to clarify the meaning of each one of them, we present a brief review \cite{Wald}.

Firstly, it is physically reasonable and even desirable that the measure of the energy density, $\rho$, of a matter field, 
a fluid, be non-negative. Thus, as the first condition we can expect that the energy density measured by an observer with 
quadrivelocity $\xi^{\mu}$ be always positive, which can be mathematically represented by
\bq
\lb{teta2}
{T}_{\mu \nu} \; {\xi}^{\mu} \; {\xi}^{\nu} \geq 0,
\eq
for all vector $\xi^{\mu}$ time-like or null. The energy condition (\ref{teta2})
is called as weak energy condition.
On the other hand, a fundamentally important equation in the study of the process of the gravitational collapse is the Raychaudhuri 
equation, given by
\bq
\lb{teta3}
{\xi}^{\mu} \nabla_{\nu} \theta = \frac{d \theta}{d \tau} = - \frac{1}{3} {\theta}^2
- \sigma^{\mu \nu} \sigma_{\mu \nu} + \omega^{\mu \nu} \omega_{\mu \nu} - {R}_{\mu \nu} {\xi}^{\mu} {\xi}^{\nu},
\eq
where $\tau$ is the proper time, $\theta$ define the expansion, 
$\sigma^{\mu \nu}$ the shear and $\omega^{\mu \nu}$ the torsion.
Equation (\ref{teta3}) describe the rate of the expansion variation through the geodesic curves in the family 
of the congruence considered. The term ${R}_{\mu \nu} {\xi}^{\mu} {\xi}^{\nu}$ is associated with the field 
matter by the equation 
\bq
\lb{teta4}
{R}_{\mu \nu} {\xi}^{\mu} {\xi}^{\nu} = \left( T_{\mu \nu} - \frac{1}{2}T  \; g_{\mu \nu} \right)
\xi^{\mu} \xi^{\nu} = {T}_{\mu \nu} {\xi}^{\mu} {\xi}^{\nu} + \frac{1}{2}T,
\eq
which must be always positive in order to give a negative contribution for the variation of the expansion of the 
geodesic curves in the congruence. Mathematically this implies in the condition 
\bq
\lb{teta5}
{T}_{\mu \nu} {\xi}^{\mu} {\xi}^{\nu} \geq -\frac{1}{2}T.
\eq
Then, with equation (\ref{teta5}) can be assured that all the matter physically well behaviored,
exerts a convergency effect on the time-like or time-null geodesic families,
guaranteeing the attractiveness of gravity.
This energy condition is known as the strong condition. It can also be called 
the convergency condition in a collapse process.

Finally, another energy condition, called as dominant condition,
imposes that the local velocity of the matter flux is always smaller than the local light velocity, 
which is guaranteed if the pressure of the fluid do not exceed the energy density. For an observer with quadrivelocity $\xi^{\mu}$,
the quantity $-{T_\nu}^\mu \; \xi^{\nu}$ physically restrict the quadrivelocity of the matter energy current, under  point 
of view of this observer. In this way, all future time-like vector $\xi^{\mu}$, $-{T_\nu}^{\mu} \; \xi^{\nu}$ must be a 
time-like or null vector.
Note that the strong energy conditions do not include the weak energy conditions. The strong concept is based only on the fact that 
the condition described in (\ref{teta5}) is physically more restrictive as that presented in equation (\ref{teta2}).

Applying these conditions to a general spherically symmetric anisotropic fluid they reduct to

\noi i) the weak energy conditions (\ref{teta2})
\bq
\rho \ge 0,
\lb{WEC1}
\eq
\bq
\rho + p_r \ge 0,
\lb{WEC2}
\eq
\bq
\rho + p_t \ge 0,
\lb{WEC3}
\eq

where $\rho$ is the energy density, $p_r$ and 
$p_t$ are the radial and tangential pressure, respectively.

\noi ii)
The strong energy conditions (\ref{teta5}) are now given by the equations 
(\ref{WEC2})-(\ref{WEC3}) and
\bq
\rho + p_r + 2p_t \ge 0.
\lb{SEC1}
\eq

\noi The dominant energy conditions are given by the equations
(\ref{WEC1})-(\ref{WEC3}) and
\bq
\rho - p_r \ge 0,
\lb{DEC1}
\eq
\bq
\rho - p_t \ge 0.
\lb{DEC2}
\eq

In the particular case of the homogeneous e isotropic 
Friedmann cosmological models \( (p_r= p_t =p) \) assume an
equation of state in the form 
\bq
\lb{seiso}
p=\omega\rho,
\eq
where $p$ is the isotropic pressure. 
In order to have accelerated expansion scenario coherent with supernovae observations,
we need to have
\bq
\lb{se}
\omega<-\frac{1}{3}.
\eq
But, in this range, the strong energy  condition (\ref{SEC1}), that reduces to
\bq
\lb{isec}
\rho + 3 p \geq 0,
\eq
\noi is violated.
This means that for an accelerated Friedmann universe is necessary to suppose a strange kind of energy,
permeating all the spacetime that was called dark energy.

Then, the denomination dark energy is applied to fluids which
violate only the strong energy condition given by the equation (\ref{SEC1}).
On the other hand, the denomination phantom energy is associated to fluids
which, besides violation of the equation (\ref{SEC1}), also violate at least
one of the conditions given by equations (\ref{WEC2}) or (\ref{WEC3}). 
More specifically we assume, to this case, the denomination repulsive phantom energy, 
while the case where only the conditions given by equation (\ref{WEC2}) or 
(\ref{WEC3}) is violated as attractive phantom energy.

On the other hand, generally, physical relevant anisotropic fluids present a state equation as 
$p_r=\omega\rho$, where $p_r$ is the radial component of the pressure and the tangential component is 
furnished by the field equations. Note that, even in this case, the inequality (\ref{WEC2}) implies 
the same condition $\omega<-1$. However, the condition on the tangential 
component (\ref{WEC3}) does not imply, in an obvious way, a dependence on the 
parameter $\omega$, since both components are independent each other. Then, if we consider the limit $\omega<-1$, in fact, 
we can identify some solutions of 
phantom fluids, but we may not see other possibilities which arise from the tangential component of the pressure.
Differently of the phantom case, the definition of the dark energy fluids, based on the violation of the condition (\ref{SEC1}), 
can take us to a entirely different limits from that originated in the isotropic case.
Thus $\omega<-1/3$ does not assure the characterization of dark energy in this  case. This occurs because the value 
of the limit for the parameter $\omega$ can depend on the other physical parameters introduced by the tangential component of
 the pressure. Besides this, it can also be considered anisotropic models where the equation of state is applied on the tangential 
component of the pressure ($p_t=\omega\rho$). In this case, all the above discussion is still valid, since we change the role of 
the pressure components.

\section{Example of Misclassification}

In the following we will present an example where this analysis should be done 
carefully.
Motivated by the picture of gravastars, Lobo \cite{Lobo} has proposed a model
of stars constituted by dark energy.  However, the analysis in his work is based
on the dark energy limits for the parameter $\omega$ of the equation of state 
$p=\omega \rho$ with
$-1 \le \omega \le -1/3$.  This interval of values comes from the Friedmann
cosmological models which assume isotropic pressures $p$.
Since the Lobo's stellar models \cite{Lobo} use anisotropic pressure fluids, it is 
necessary a revision of the interval of $\omega$, in order to have a correct
classification of dark, not
dark (or standard) and phantom energy. Moreover, for the fluid proposed in 
his work to represent the interior of the star, it is necessary to introduce a 
subclassification in order to describe all the different behavior of the 
gravity.

Applying the energy conditions to the first Lobo's model \cite{Lobo} where
\bq
\rho=\rho_0,
\eq
\bq
p_r=\omega \rho_0,
\eq
\bq
p_t=\omega \rho_0 \left[ 1 + \frac{4\pi}{6} \frac{(1+\omega)(1+3\omega)\rho_0 r^2}
{\omega \left(1- \frac{8\pi \rho_0}{3}r^2 \right)} \right],
\eq
{\small the conditions} from equations (\ref{WEC2}) and (\ref{WEC3}) give us
\bq
\omega\ge -1,
\eq
and
\bq
\label{10}
3m(r)\omega^2 + 2\omega r + 2r -3m(r) \ge 0,
\eq
where $m(r)=4\pi\rho_0 r^3/3$ and $r$ are the mass and the radial coordinate, 
respectively. Observe that the contribution of the tangential pressure introduce a dependence 
between $\omega$, $m(r)$ and $r$, which was forgot in the Lobo's work
\footnote{We would like to stress out that in a recent comment \cite{Lobo08} of a previous
version of this paper \cite{Chan08b} , Lobo has suggested that we have considered $p_r=\omega \rho$
and $p_t=\omega \rho$ as the equation of state.  This fact is obviously incorrect
since we have used his original anisotropic equation of state, as an example \cite{Lobo}.}.

Here we present the analysis only of the first Lobo's model, as
an example, by virtue of its simplicity.

The condition given by equation (\ref{10}) is satisfied in the following intervals
\begin{enumerate}
\item for $r=3m(r)$ for any values of $\omega$,
\item for $2m(r) < r < 3m(r)$, we have $\omega < -1$ and $\omega > 1-2r/[3m(r)]$,
\item for $r > 3m(r)$, we have $\omega < 1-2r/[3m(r)]$ and $\omega > -1$.
\end{enumerate}
On the other hand, the strong energy condition, from equation (\ref{SEC1}), furnishes us
\bq
3m(r) \omega^2 + [3r - 2m(r)]\omega + r - m(r) \ge 0.
\eq
Thus dark energy is possible for the range $1-r/m(r)<\omega<-1/3$. 
This fact implies that some of the solutions considered by Lobo as being dark
energy fluids ($\omega<1-r/m(r)$), which we denote here as attractive phantom energy, 
are misclassified
\footnote{This equation is correct although the Lobo's comment \cite{Lobo08}
 claims be incorrect.}.

We have summarized the limits for $\omega$ in figure 1, which depend on the 
values of $m(r)$ and $r$. Note that in the particular cases where the Lobo's 
solution corresponds to a isotropic pressure fluids ($\omega=-1$ and 
$\omega=-1/3$) his classification for fluids is preserved.  However,
the frontier value of $\omega$ for which the fluid pass from dark energy to 
repulsive phantom energy, in the anisotropic case, it only coincides with
the value $\omega=-1$ for $r\ge 3m(r)$. 
Note that for the range $2m(r)<r<3m(r)$, the limit between dark 
energy and repulsive phantom energy is given by the violation of the energy 
condition $\rho+p_t\ge 0$, corresponding to the contribution of the pressure 
which is not defined by an equation of state, instead of the condition 
$\rho+p_r\ge 0$. This is sufficient to modify the limit value of $\omega$, 
where the 
fluid is constituted by repulsive phantom energy, from $-1$ to $1-2r/[3m(r)]$.
Then, for this simple example treated here, we have a clear evidence that this 
limit depends strongly on the rate between $r$ and $m(r)$. A relationship 
between the magnitude of anisotropy and the local compactness for 
gravastars was already 
suggested by Cattoen, Faber and Visser \cite{Cattoen}. Moreover we can also 
see that there is a 
new and significant frontier given by $\omega=1-r/m(r)$, for all ratio
$r/m(r)$.

\myfigure{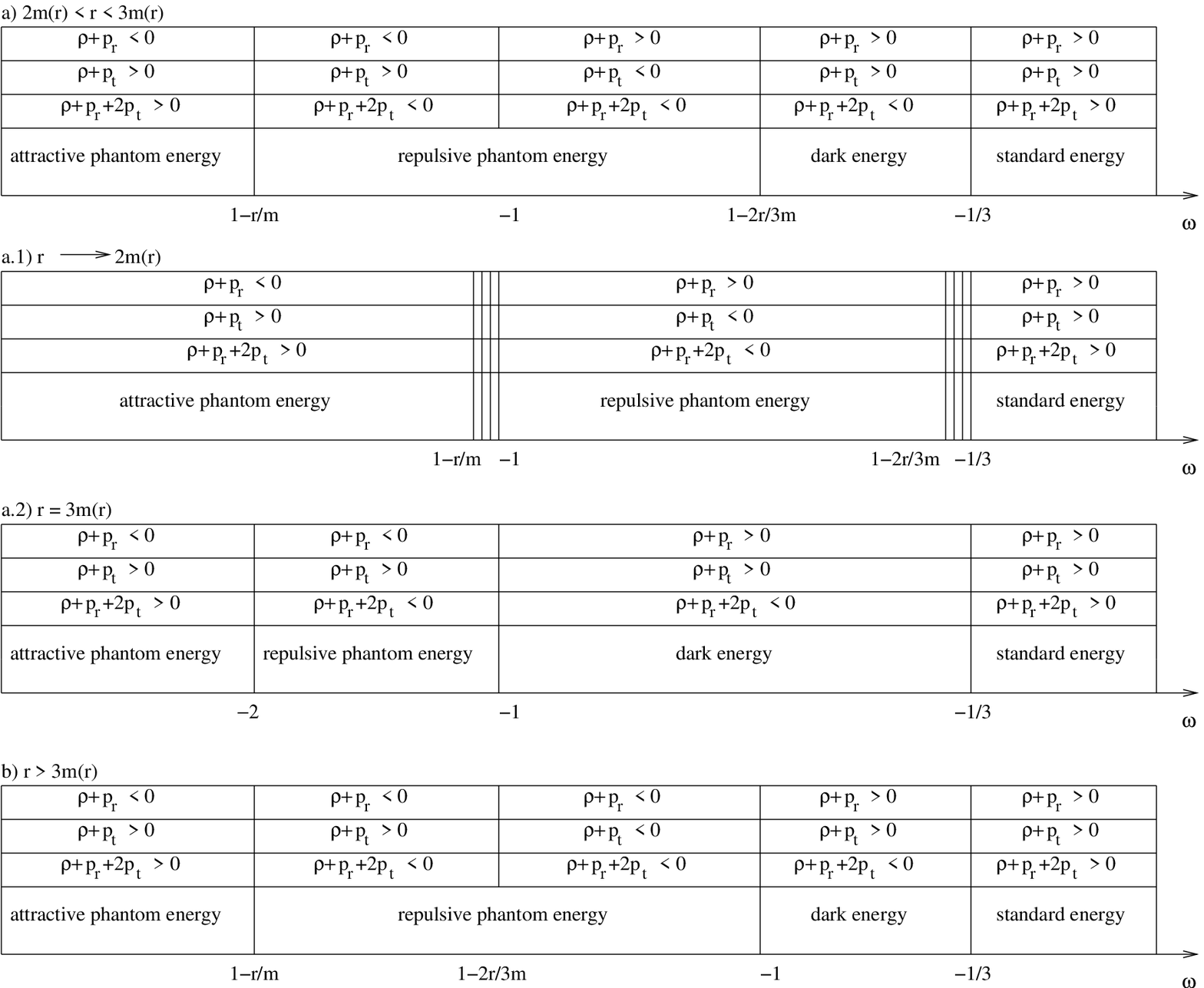}{Dependence of the fluid's nature on $\omega$.
The several parallel lines in the case (a.1) denote that exist regions
where still there is repulsive phantom energy ($1-r/m(r) < \omega < -1$) and
dark energy ($1-2r/[3m(r)] < \omega < -1/3$).}{figure1}

\section{Conclusions}

In this work we have shown that the definition of the dark energy and 
phantom fluids, in terms of the energy conditions, must be carefully used
for systems
where anisotropy may be very important for gravitational collapse.
In particular, we point out that the pressure components may have very
important roles and they can contribute differently for the classification
of the fluid and its gravitation behavior (attractive or repulsive).
In this work we present an example where the
classification of dark energy for isotropic pressure fluids is used
incorrectly for anisotropic fluids.  The correct classification and its
consequences has been shown.
 
Finally, we would like to stress that the correction proposed here does not 
invalidate the results and conclusions presented by Lobo \cite{Lobo}, although 
it restricts his solutions which characterize dark energy stars. We consider
 that the care pointed out here, and exemplified by us through one of the 
Lobo's solutions, with anisotropic pressure fluids is necessary and must be 
observed in any work with anisotropic pressure models (star or cosmological 
ones).

\section*{Acknowledgments}

The financial assistance from CNPq (JFVR) and  
FAPERJ/UERJ (MFAdaS) are gratefully acknowledged. The
author (RC) acknowledges the financial support from FAPERJ (no.
E-26/171.754/2000, E-26/171.533/2002 and E-26/170.951/2006).  
The authors (RC and MFAdaS) also acknowledge the financial support from 
Conselho Nacional de Desenvolvimento Cient\'{\i}fico e Tecnol\'ogico - 
CNPq - Brazil.  The author (MFAdaS) acknowledges the financial support
from Financiadora de Estudos e Projetos - FINEP - Brazil (Ref. 2399/03).


\section*{References}

\begin{thebibliography}{100}
\bibitem{agr98} A.G. Riess  {\it et al.}, Astron. J. {\bf 116}, 1009 (1998).

\bibitem{agr98a} S. Perlmutter {\it et al.}, Astrophys. J. {\bf 517}, 565 (1999).

\bibitem{Obs1} A.G. Riess {\it et al.}, Astrophys. J. {\bf 607}, 665 (2004).

\bibitem{Obs2} P. Astier {\it et al.}, Astron. and Astrophys. {\bf 447}, 31 (2006).

\bibitem{Obs3} D.N. Spergel {\it et al.}, astro-ph/0603449.

\bibitem{Obs4} W.M. Wood-Vasey {\it et al.}, astro-ph/0701041.

\bibitem{Obs5} T.M. Davis {\it et al.}, astro-ph/0701510.

\bibitem{wmap}C.~L.~Bennett {\it et al.}, Astrophys.\ J.\ Suppl.\  {\bf 148},
1 (2003) [arXiv:astro-ph/0302207].

\bibitem{wmap1} D.~N.~Spergel {\it et al.}  [WMAP Collaboration],
Astrophys.\ J.\ Suppl.\  {\bf 148}, 175 (2003) [arXiv:astro-ph/0302209].

\bibitem{sdss}M.~Tegmark {\it et al.}  [SDSS Collaboration],
Phys.\ Rev.\ D {\bf 69}, 103501 (2004) [arXiv:astro-ph/0310723].

\bibitem{sdss1} K.~Abazajian {\it et al.} (2004) [arXiv:astro-ph/0410239].

\bibitem{sdss2} K.~Abazajian {\it et al.} [SDSS Collaboration],
Astron.\ J.\  {\bf 128}, 502 (2004) [arXiv:astro-ph/0403325].

\bibitem{sdss3} K.~Abazajian {\it et al.} [SDSS Collaboration],
Astron.\ J.\  {\bf 126}, 2081 (2003) [arXiv:astro-ph/0305492].

\bibitem{sdss4} E.~Hawkins {\it et al.},
Mon.\ Not.\ Roy.\ Astron.\ Soc.\  {\bf 346}, 78 (2003)
[arXiv:astro-ph/0212375].

\bibitem{sdss5} L.~Verde {\it et al.},
Mon.\ Not.\ Roy.\ Astron.\ Soc.\  {\bf 335}, 432 (2002)
[arXiv:astro-ph/0112161].

\bibitem{quint1} R.~R.~Caldwell, R.~Dave and P.~J.~Steinhardt, Phys. Rev. Lett. {\bf 80}, 1582 ( 1998).

\bibitem{quint2} A.~R.~Liddle and R.~J.~Scherrer, Phys. Rev. D {\bf 59}, 023509 (1999).

\bibitem{quint3} P.~J.~Steinhardt, L.~M.~Wang and I.~Zlatev, Phys. Rev. D {\bf 59}, 123504 (1999).

\bibitem{DGP1} G.~R.~Dvali, G.~Gabadadze and M.~Porrati, Phys. Lett. B {\bf 484}, 112 (2000).

\bibitem{DGP2} C.~Deffayet, Phys. Lett. B {\bf 502}, 199 (2001).

\bibitem{FR1} S. Capozziello, S. Carloni, and A. Troisi, (2003) [arXiv:astro-ph/0303041].

\bibitem{FR2} S.M. Carroll, et al, Phys. Rev. D{\bf 70}, 043528 (2003).

\bibitem{FR3} S. Nojiri and S.D. Odintsov, Phys. Rev. D {\bf 68}, 123512 (2003).

\bibitem{other1} P.K. Townsend and N.R. Wohlfarth, Phys. Rev. Lett. {\bf 91}, 061302 (2003).

\bibitem{other2} G. W.Gibbons in {\it Supersymmetry, Supergravity and Related Topics},
\bibitem{other3} J.M. Maldacena  and C. Nu\~{n}ez, Int. J. Mod. Phys. A{\bf 16}, 822 (2001).

\bibitem{other4} N. Ohta, Phys. Rev. Lett. {\bf 91}, 061303 (2003).

\bibitem{other5} N.R. Wohlfarth, Phys. Lett. B {\bf 563}, 1 (2003).

\bibitem{other6} S. Roy, Phys. Lett. B {\bf 567}, 322 (2003).

\bibitem{other7} J.K. Webb,  {\it et al},  Phys. Rev. Lett. {\bf 87}, 091301 (2001).

\bibitem{other8} J.M. Cline and J. Vinet, Phys. Rev. D{\bf 68}, 025015 (2003).

\bibitem{other9} N. Ohta,  Prog. Theor. Phys. {\bf 110}, 269 (2003).

\bibitem{other10} N. Ohta, Int. J. Mod. Phys. A {\bf 20}, 1 (2005).

\bibitem{other11} C.M. Chen, {\it et al}, JHEP, {\bf 10}, 058 (2003).

\bibitem{other12} E. Bergshoeff, Class. Quantum Grav. {\bf 21}, 1947 (2004).

\bibitem{other13} Y. Gong and A. Wang, Class. Quantum Grav. {\bf 23}, 3419 (2006).

\bibitem{other14} I.P. Neupane and D.L. Wiltshire,  Phys. Lett. B {\bf 619}, 201 (2005).

\bibitem{other15} I.P. Neupane and D.L. Wiltshire, Phys. Rev. D {\bf 72}, 083509 (2005).

\bibitem{other16} K. Maeda and N. Ohta, Phys. Rev. D, {\bf 71}, 063520 (2005).

\bibitem{other17} I.P. Neupane, Phys. Rev. Lett. {\bf 98}, 061301 (2007).

\bibitem{other18} Y. Gong, A. Wang and Q. Wu, (2007) [arXiv:gr-qc/0711.1597].

\bibitem{Pereira06} P.R.C.T. Pereira, M.F.A. da Silva and R. Chan,
IJMPD {\bf 15}, 991 (2006).

\bibitem{Brandt07} C.F.C. Brandt, R. Chan, M.F.A. da Silva, J.F. Villas da Rocha, Gen. Relat. Grav. {\bf 39}, 1675 (2007).

\bibitem{DE1} E.J. Copeland, M. Sami and S. Tsujikawa, Int. J. Mod. Phys. D {\bf 15}, 1753 (2006).

\bibitem{DE2} T. Padmanabhan, (2007) [arXiv:gr-qc/0705.2533].

\bibitem{Koivisto} T. Koivisto and D.F. Mota, J. Cosmol. Astropart. Phys. {\bf 06}, 018 (2008).

\bibitem{Ma99} C. Ma, R.R. Caldwell, P. Bode, and L. Wang, Astrophys. J.
{\bf 521}, L1 (1999).

\bibitem{FJ97} P.G. Ferreira and M. Joyce, Phys. Rev. Lett. {\bf 79},
 4740 (1997).

\bibitem{FJ97a} D.F. Mota, C. van de Bruck, Astron. Astrophys.
{\bf 421}, 71 (2004).

\bibitem{Mazur02} Mazur, P. O. and Mottola, E. (2002) [arXiv:gr-qc/0109035] .

\bibitem{Visser04} Visser, M and Wiltshire, D. L., Class. Quantum Grav. {\bf 21}, 1135 (2004).

\bibitem{DEStar} http://www.nature.com/news/2005/050328/full/050328-8.html.

\bibitem{DEStar1} http://www.bioon.com/TILS/news/200504/97002.html.

\bibitem{Chan08a} 
Rocha, P., Miguelote, A. Y., Chan, R., da Silva, M. F., Santos, N. O. and Wang, A.,
J. Cosmol. Astropart. Phys. {\bf 06}, 025 (2008).

\bibitem{BDE04} E. Babichev, V. Dokuchaev, and Yu. Eroshenko, Phys. Rev.
Lett. {\bf 93}, 021102 (2004).

\bibitem{BDE04a} S. Nojiri, S.D. Odintsov, Phys. Rev.
{\bf D70}, 103522 (2004) [arXiv:hep-th/0408170].

\bibitem{BHDE1} Z.-H. Li, A. Wang, Mod. Phys. Lett. A22, 1663-1676 (2007) [arXiv:astro-ph/0607554].

\bibitem{BHDE2} R.-G. Cai and A. Wang, Phys. Rev. D73, 063005 (2006) [arXiv:astro-ph/0505136].

\bibitem{BHDE3} S. Nath, S. Chakraborty, and U. Debnath, (2005) [arXiv:gr-qc/0512120].

\bibitem{BHDE4} U. Debnath and S. Chakraborty, (2006) [arXiv:gr-qc/0601049].

\bibitem{Paramos} Bertolami, O., P\'aramos, J., Phys. Rev. D {\bf 72}, 123512 (2005) 
[arXiv:astro-ph/0509547]

\bibitem{Lobo07} Lobo, F. (2007) [arXiv:gr-qc/0611083].

\bibitem{CattoenVisser05} Cattoen, C., Faber, T. and Visser, M. Class. Quantum Grav. {\bf 22}
4189 (2005).

\bibitem{Lobo} Lobo, F., Class. Quant. Grav. {\bf 23}, 1525 (2006).

\bibitem{Chan08} Chan, R., da Silva, M.F.A., Villas da Rocha, J.F. (2008) [arXiv:gr-qc/0803.3064].

\bibitem{Ruderman} Ruderman, M., Ann. Rev. Astron. Astrophys. {\bf 10}, 427 (1972).

\bibitem{Canuto} Canuto, V., Neutron Stars: General Review Proc. Solvay Conf. on Astrophysics and Gravitation (Brussels, 1973).

\bibitem{Hartle} Hartle, J. B., Sawyer, R. F. and Scalapino, D. J., Astrophys. J. {\bf 99}, 471 (1975).

\bibitem{Itoh} Itoh, N., Prog. Theor. Phys. {\bf 44}, 291 (1970).

\bibitem{Sokolov} Sokolov, A.I.,  Sov. Phys.-JETP {\bf 52}, 575 (1980).

\bibitem{Ruffini} Ruffini, R. and Bonazzola, S., Phys. Rev. {\bf 187}, 1767 (1969).

\bibitem{Gleiser} Gleiser, M., Phys. Rev. D {\bf 38}, 2376 (1988).

\bibitem{Kippenhahn} Kippenhahn, R. and Weigert, A., Stellar Structure and Evolution (Berlin: Springer), p. 255 (1990).

\bibitem{Letelier} Letelier, P., Phys. Rev. D {\bf 22}, 807 (1980).

\bibitem{Bayin} Bayin, S., Phys. Rev. D {\bf 26}, 1262 (1982).

\bibitem{Ralston} Ralston, I. and Smith, L., Astrophys. J. {\bf 367}, 54 (1991).

\bibitem{Madsen} Madsen, I. Astrophys. J. {\bf 367}, 507 (1991).

\bibitem{Bowers} Bowers, R.I. and Liang, E.P.T., Astrophys. J. {\bf 188}, 657 (1974).

\bibitem{Herrera} Herrera, L. and Nunez, L.A., Astrophys. J. {\bf 339}, 339 (1989).

\bibitem{Chan1} Chan, R., Herrera, L. and Santos, N.O., Class. Quantum Grav. {\bf 9}, L133 (1992).

\bibitem{Chan2} Chan, R., Herrera, L. and Santos, N.O., Mon. Not. R. Astron. Soc. {\bf 265}, 533 (1993).

\bibitem{Abreu} Abreu, H., Hernandez, H., Nunez, L. A., Class. Quantum Grav. {\bf 24}, 4631 (2007).

\bibitem{Cattoen} Cattoen, C., Faber, T. and Visser, M. Class. Quantum grav. {\bf 22} 4189 (2005).

\bibitem{Wald} Wald, R. M. {\em General Relativity}, (University of Chicago 
Press, Chicago and London, 1984, p. 218).

\bibitem{Lobo08} Lobo, F. (2008) [arXiv:gr-qc/0805.2309].

\bibitem{Chan08b} Chan, R., da Silva, M.F.A., Villas da Rocha, J.F. (2008) [arXiv:gr-qc/0803.2508].

\end{thebibliography}
\end{document}